\documentclass[aps,prd,twocolumn]{revtex4-2}
\usepackage{graphicx,epsfig}
\usepackage{amsmath,amssymb}
\usepackage{lineno,hyperref}
\hypersetup{colorlinks=true}
\usepackage{xcolor}

\usepackage{caption}
\usepackage{comment}
\newcommand{\ii}{\mathrm{i}} 

\renewcommand{\phi}{\varphi}

\hypersetup{
 colorlinks=true,
 linktoc=all,
 linkcolor=red,
 citecolor=blue,
 urlcolor = blue}

\begin{document}
\title{Compact Binaries through a Lens: Silent vs.\ Detectable Microlensing for the LIGO-Virgo-KAGRA Gravitational Wave Observatories}
\begin{abstract}
     Massive objects located between Earth and a compact binary merger can act as gravitational 
lenses magnifying signals and improving the sensitivity of gravitational wave detectors to distant events. 
Depending on the parameters of the system, a point mass lens between the detector
and the source can either lead to a smooth frequency-dependent amplification of the
gravitational wave signal, or magnification combined with the appearance of a second
image that interferes with the first creating a regular, predictable pattern.
     We map the increase in the signal to noise ratio for upcoming LVK observations as a function of the mass of the lens $M_L$ and dimensionless source position $y$ for any point mass lens between the detector and the binary source.   To quantify detectability, we compute the optimal match between the lensed waveform and the waveforms in the unlensed template bank and provide a map of the match. The higher the mismatch with unlensed templates, the more detectable lensing is. 
       Furthermore, we estimate the probability of lensing, and find that the redshift to which binary mergers are visible with the LVK increases from $z \approx 1$ to $z\approx 3.2$ for a total detected mass $M_{det} = 120 M_\odot$. The overall probability of lensing is $<20\%$ of all detectable events above the threshold SNR for $M_{det} = 120 M_\odot$ and $<5\%$ for more common events with $M_{det} = 60 M_\odot$.   We find that there is a selection bias for detectable lensing that favors events that are close to the line of sight $y  \lesssim 0.5$. Black hole binary searches could thus improve their sensitivity by taking this bias into account.   Moreover, the match, the SNR increase due to lensing, and the probability of lensing are only weakly dependent on the noise curve of the detector with very similar results for both the O3 and predicted O4 noise power spectral densities. These results are upper limits that assume all dark matter is composed of $300 M_\odot$ point mass lenses.
\end{abstract}

\author {Ruxandra Bondarescu}
\email{ruxandra@icc.ub.edu}
\affiliation{Institut de Ci\`encies del Cosmos (ICCUB),
Facultat de F\'{i}sica,
 Universitat de Barcelona, Mart\'i i Franqu\`es 1, E-08028 Barcelona, Spain}
\author{Helena Ubach}
 \email{helenaubach@icc.ub.edu}
 \affiliation{Institut de Ci\`encies del Cosmos (ICCUB),
Facultat de F\'{i}sica,
 Universitat de Barcelona, Mart\'i i Franqu\`es 1, E-08028 Barcelona, Spain}
 \author{
Andrew Lundgren}
\email{andrew.lundgren@port.ac.uk}
\affiliation{Institute of Cosmology and Gravitation, University of Portsmouth, Dennis
Sciama Building, Burnaby Road, Portsmouth, PO1 3FX, United Kingdom}
\author{Oleg Bulashenko}
\email{oleg@fqa.ub.edu}
\affiliation{Institut de Ci\`encies del Cosmos (ICCUB),
Facultat de F\'{i}sica,
 Universitat de Barcelona, Mart\'i i Franqu\`es 1, E-08028 Barcelona, Spain}

\date{\today}

\maketitle

\section{Introduction}
Gravitational wave detectors register oscillations in the fabric of space-time, which propagate at the speed of light. These waves are not absorbed by intervening matter providing a unique insight into the dynamics of the universe.  Since the spacetime is very stiff, gravitational waves are very weak, and can only be detected on Earth if they come from the inspiral and merging of very dense compact objects like binary black holes or neutron stars. Black holes have the strongest gravitational wave emission and can be detected further away than other compact objects. This makes them more likely candidates for gravitational lensing, where an object close to the line of sight interposes between the source and the detector amplifying the gravitational wave emission. While black holes are fully characterized by mass, spin and charge, the emitted gravitational waves from a binary merger represent the last act in a million or billion year chain of events. 
Reconstructing the characteristics of the binary in the source frame where the collision happened is the primary challenge after the identification of the signal.  

  The first merging black holes detected from the emitted gravitational radiation were unlike any black holes seen in the Milky Way before. They are quite massive, substantially more than black holes found in X-ray studies, and appear not to have much spin.  Since the first detection in 2015 \cite{LIGO16-1}, the LIGO-Virgo-Kagra (LVK) collaboration has  registered about $90$ gravitational wave events \cite{LIGO18-GWTC1, LIGO20-GWTC2, LIGO21-GWTC3}. The average total mass for detected gravitational wave binaries has been observed to be $\approx 60 M_\odot$, which is unexpectedly high. One proposed explanation for such high masses is gravitational lensing \citep{diego19, diego20}.


While gravitational wave signals provide information that is complementary to telescope observations, they are challenging to interpret correctly. Massive objects within an Einstein cone of the detected compact object collision alter the signal causing gravitational lensing \cite{schneider-92, petters-01}.
It can result in magnified signals, duplicate events, or conspire to create a beating pattern in the waveform which occurs when the time delay between the paths is comparable with the period of the wave. Repeated events separated by minutes to months occur when the lens is a galaxy \cite{sereno10,ng18,li18,oguri18,oguri19}. The separation can increase up to years for galaxy clusters \cite{smith18,ryczanowski20,robertson20}. Here we focus on microlensing by stars or other compact objects (duplicate events separated by milliseconds to seconds). Microlensing in the context of gravitational waves has been the focus of a plethora of studies \cite{nakamura98,nakamura99,takahashi03,matsunaga06,cao14,lai18,christian18,liao19,
diego19,hou20,orazio20,diego20,cheung21,hou21,cremonese21,cremonese21b,yu21,biesiada21,suvorov21,chung21,dalang22,BU-22,caliskan22,tambalo22}. 

Careful modeling can turn gravitational lensing from a hindrance in waveform recovery to a new instrument in gravitational wave astrophysics.
Some gravitational lenses could be objects that do not emit light or neutrinos and cannot be observed through other channels. Lensing could then constrain populations of unknown or poorly understood objects like intermediate mass black holes \cite{lai18,gais22}, dark matter haloes \cite{choi21,gao22,guo22} or topological defects \cite{suyama06,pla-string16,pla-fresnel17} and even prove the existence of new objects. When we compute the number of lenses we take the amount of dark matter in the universe split into microlenses as an upper limit. A first detection of gravitational lensing could provide the first direct evidence of compact dark matter objects \cite{jung19,liu19,liao20,choi21,wang21,urrutia22,gao22,basak22,fairbairn22,tambalo22b}.
 
The lensing amplification causes the source's distance from the detector to be underestimated. Not only does it appear to be closer, but it also appears more massive because the redshift is underestimated. It is thus crucial to determine whether a detected gravitational wave signal is lensed or not, because misidentification results in wrong parameter estimation, e.g., if the signal is magnified due to the presence of massive objects in between the binary inspiral and Earth it will appear further than it actually is. A binary at $z=3$ will appear four times more massive when detected in the detector frame than in the source frame with $M_{\rm det} = (1+z) M_{\rm src}$.  A population of objects with suspicious distances or masses could provide indirect evidence of lensing. {\it In this paper we ask the questions of when microlensing of an individual source can be clearly detected, and how would such a detection appear}. Characteristics of lensing include (1) the appearance of a second image, (2) a regular pattern of minima and maxima that is predictable or simply (3) an amplification of the signal that increases with frequency and is maximum at merger. 




While the LVK collaboration analyzed O2 and O3 observation runs \cite{hannuksela19,lensingO3a, lensingO3b}, they did not find confident evidence for gravitational-wave lensing.  The detection rate is expected to increase from one gravitational wave per week in O3 to a possible one per day in O4 \cite{detectors20, observing-plan}. The higher number of sources elevate the probability of detecting a detectably lensed event.  Furthermore, plans for new, advanced detector facilities are crystalizing in both Europe via the Einstein Telescope and the US via the Cosmic Explorer \cite{auger-plagnol-17, maggiore-20, barrause-20, bailes-21}. 

Here we investigate scenarios for detection of microlensing of binary black hole coalescences. We make three key contributions. We provide simple but accurate approximations for the SNR and the maximum mismatch between microlensed waveforms and an unlensed template bank which determines when lensing is detectable rather than silent.  We show the SNR increase and mismatch are largely independent of the BBH masses or the detector's sensitivity curve for current-generation detectors. Moreover, we provide the first derivation of an important selection bias which affects parameter estimation.

Sec. \ref{sec_pml} begins by reviewing the effect of a point-mass lens (PML) on the gravitational-wave signal. The point-mass lens is the simplest most illustrative model for a microlensing object. The lensing effects depend only on the lens mass $M_L$ and $y$, which gives the position of the lens relative to the line of sight between the source and the detector. We compute the relative increase in the signal-to-noise ratio (SNR) caused by the lensing amplification of the signal, which induces detectability to larger distances and show that it is similar for O3 and O4 noise curves. The parameter space in divided in three regions: the amplification region (the SNR increase depends only on $M_L$, {   see Appendix A}), a transition region  (the SNR increase depends on both $y$ and $M_L$), and the geometrical optics region (the oscillations in the transmission factor average out, and the SNR increase depends only on $y$, {   Appendix A}).  

In Sec. \ref{match} the mismatch of the lensed signal versus unlensed signals is calculated. We quantify the detectability by the mismatch of the lensed signal versus any unlensed templates. This is the fundamental ingredient in model comparisons since without sufficient mismatch the signal intrinsically does not contain information to indicate that it is lensed. As we were submitting this manuscript,  Ali {\it et al} \cite{ali-23} published the mismatch for strong lensing in the Point Mass Lens (PML) and the single isothermal sphere cases. They propose model-independent lensing parameters. Their analysis suggests detectable values at $y\approx 0.7$ for $M_L = 10^3 M_\odot$ for the PML, which is consistent with ours. However, the paper does not go beyond the mismatch calculation to estimate the probability of lensing.

 Crucially, Sec \ref{rates} evaluates not just the rate of lensing, but the rate of \emph{detectable} lensing.
Otherwise, lensing is \emph{silent}. In silent lensing, the distance to the binary and the mass of the binary could be affected without creating a detectable mismatch with unlensed templates.
Selecting only for lensed waveforms that carry enough of a lensing imprint to be measured limits parameter space improving prior distribution of lens parameters for detecting microlensing.  

Simulating microlensing, or doing Bayesian parameter estimation of possibly lensed events, requires prior distributions for the lens mass $M_L$ and the dimensionless lens angle $y$. The obvious choice for priors is to model the distribution of lenses expected to exist in the universe. For the angle this is simple because randomly distributed lenses will follow a scale-free distribution $p(y) \propto y dy$ regardless of lens mass. However, this puts the most weight on lenses farthest from the line of sight, which have a negligible effect on the signal. Even when an artificial limit is put on $y$, the prior most heavily weights lenses at the boundary, which for $y > 1$ are systems that are not meaningfully lensed.

 We demonstrate the existence of a selection bias, which  reshapes the distribution of the lensing parameters when conditioning on detectability. A source must be detected above a given SNR threshold in order to be analyzed. Lenses closer to the line of sight cause more amplification, increasing the sensitive volume and greatly enhancing their relative rate of detections. Hence the distribution of observable lensed systems is peaked toward smaller lens angle. The distance to the furthest source is also much higher than typical for unlensed systems. These effects must be accounted for in simulations, and they also shape our expectations for what lensed systems are most likely to be detected.

We note that the effect of lensing depends on the redshifted lens mass, while the most natural prior gives our belief about the mass of the lens in its own frame. Hence there is an extra variable, the lens distance, which must be tracked. Our approach naturally incorporates this as well. However, in this first paper, we do not include a specific model that predicts the number and masses of lenses since they are so uncertain. We leave that to future work.  Instead we show the priors when they are conditioned as well on the lensed waveform having a sufficient mismatch that it leaves a measurable imprint on the signal. Conclusions and future directions are presented in Sec. \ref{conclusions}.  Appendix A discusses the transmission factor in the Amplification and GO regions, while Appendix B contains the simplified formulas for SNR increase due to lensing in these regions. Appendix C presents the imprint of the factor on the waveform itself in the time and frequency domain {including typical spectrograms of lensed waveforms}.

\section{Point mass lens}
\label{sec_pml}
All massive astrophysical objects are potential lenses that can be encountered by a passing gravitational wave.  
For gravitational lensing the point mass lens model (PML) is valid when the dimension of the lens is much smaller than the Einstein radius, e.g., for black holes, dense dark matter clumps, etc. 
Due to its simplicity, the PML model has been used in the literature for interpreting both electromagnetic \cite{deguchi86a,deguchi86b,schneider-92,petters-01} and gravitational wave lensing \cite{nakamura98,nakamura99,takahashi03,matsunaga06,cao14,lai18,christian18,jung19,liao19,
hou20,orazio20,cremonese21,cremonese21b,yu21,
biesiada21,chung21,suvorov21,chung21,dalang22,BU-22,caliskan22,tambalo22,gais22,jung19,liao20,choi21,wang21,urrutia22,gao22,basak22,fairbairn22}. 

Generally, lensing effects become significant when the source, the lens and the observer are all aligned within the Einstein angle $\theta_{\rm E}=R_{\rm E}/d_L$, i.e., the lens is located near the line of sight
 -- defined as the line joining the observer and the center of the lens. The Einstein radius is given by
\begin{equation}
    R_{\rm E} = \sqrt{2{\cal R_{\rm S}}\, d_{LS}\,d_{\rm L}/d_S},
\end{equation}
which is much smaller than the angular diameter distances $d_L$, $d_S$ and $d_{LS}$ from the lens to the observer, from the source to the observer, and between the lens and the source,  respectively.  Here ${\cal R_{\rm S}} =2GM_{L}/c^2$ is Schwarzschild radius of the lens, which is directly proportional to the mass of the lens $M_L$.





\subsection{Transmission Factor}
\label{factor}

Since lensing occurs in a relatively narrow region compared to the cosmological distances traveled by the wave ($d_L$, $d_S$ and $d_{LS}$), the transmission factor can be computed in the thin lens approximation.  The lens mass can then be projected onto a lens plane. In this approximation, the gravitational waves propagate freely outside the lens and interact only with a two-dimensional gravitational potential at the lens plane, where their trajectory is suddenly modified via the transmission factor $F$. 
For the PML it is given by \cite{schneider-92,deguchi86a}
\begin{equation}
F = e^{\frac{1}{2}\pi^2 \nu} e^{\ii \pi \nu  \ln (\pi \nu)} \,
\Gamma ( 1- \ii \pi \nu ) \,_1F_1 ( \ii \pi \nu; \,1; \,\ii \pi \nu y^2 ) ,
\label{F_PML}
\end{equation}
where $\nu \equiv f t_M$ is the frequency of the gravitational wave $f$ scaled by the characteristic time $t_M$, $\Gamma(z)$ is the Gamma function, and $_1 F_1(a,b,z)$ is the confluent hypergeometric function.
The transmission factor represents the ratio between the lensed wave field received by the observer, and the unlensed one (what would be observed when no lens was present). Our transmission factor is the complex conjugate of that in some other works, but is appropriate to the sign convention that we use for Fourier transforms.

As seen from Eq.~\eqref{F_PML},
the transmission factor, which is a function of frequency, depends on two parameters:
\begin{itemize}
\item the mass of the lens $M_L$ through the time
\begin{equation}
t_M =2{\cal R_{\rm S}}/c \;\approx \; 2 \times 10^{-5} \,{\rm s}\; (M_L/M_\odot),
\label{tM}
\end{equation}
\item the scaled offset of the source 
\begin{equation}
y=\theta_S/\theta_\text{E},
\end{equation}
where $\theta_S$ is the angular position of the source with respect to the line of sight.

\end{itemize}
Throughout, we will measure the mass of the lens in the observer's frame. This is related to the intrinsic mass $M_{L,0}$ by $M_L = (1 + z_L) M_{L,0}$, where $z_L$ is the cosmological redshift of the lens. In Sec.~\ref{rates}, we will revert to using the intrinsic mass and including the redshift factor explicitly.

The strain of the gravitationally lensed signal $\tilde{h} (f)$ (that would finally be detected) is the product of the unlensed strain $\tilde{h}_{UL} (f)$ and the transmission factor $F (f)$ in the frequency domain:
\begin{equation}
\tilde{h} (f) = \tilde{h}_{UL} (f) \cdot F (f).
\label{lensed_strain}
\end{equation}
We are interested in the behavior of the transmission factor $F$,
shown in Fig.~\ref{fig:regions} and detailed in Appendix A,
for the frequency region the LVK is sensitive to, which begins at about 15 Hz and ends at merger for the BH binaries considered here.

For a given choice of parameters $M_L$ and $y$, the lensing effect on GWs in the LVK range 
(shown by vertical dashed lines in Fig.~\ref{fig:regions}) will be different depending on the parameters' values, as explained in Appendices A and B. The frequency range
can fall into the GO oscillating region (higher masses and/or $y$), the amplification region (lower masses and/or $y$), or into the intermediate part, 
as seen in Fig.~\ref{fig:regions}.
If the mass is very low (e.g. $M_L=30 M_\odot$), the effect will be mainly amplification. If the mass is higher, then oscillations will appear (see Appendix C for the effects on the waveform). The spacing of the oscillations will dependent on the 
product $y \, M_L$ [Eq.~\eqref{eq:freq-spacing}],
while the amplification of the maxima and minima will be only dependent on $y$ [Eq.~\eqref{F_GO_maxmin}].

\begin{figure}[!htbp]
\includegraphics[width=1.0\columnwidth]{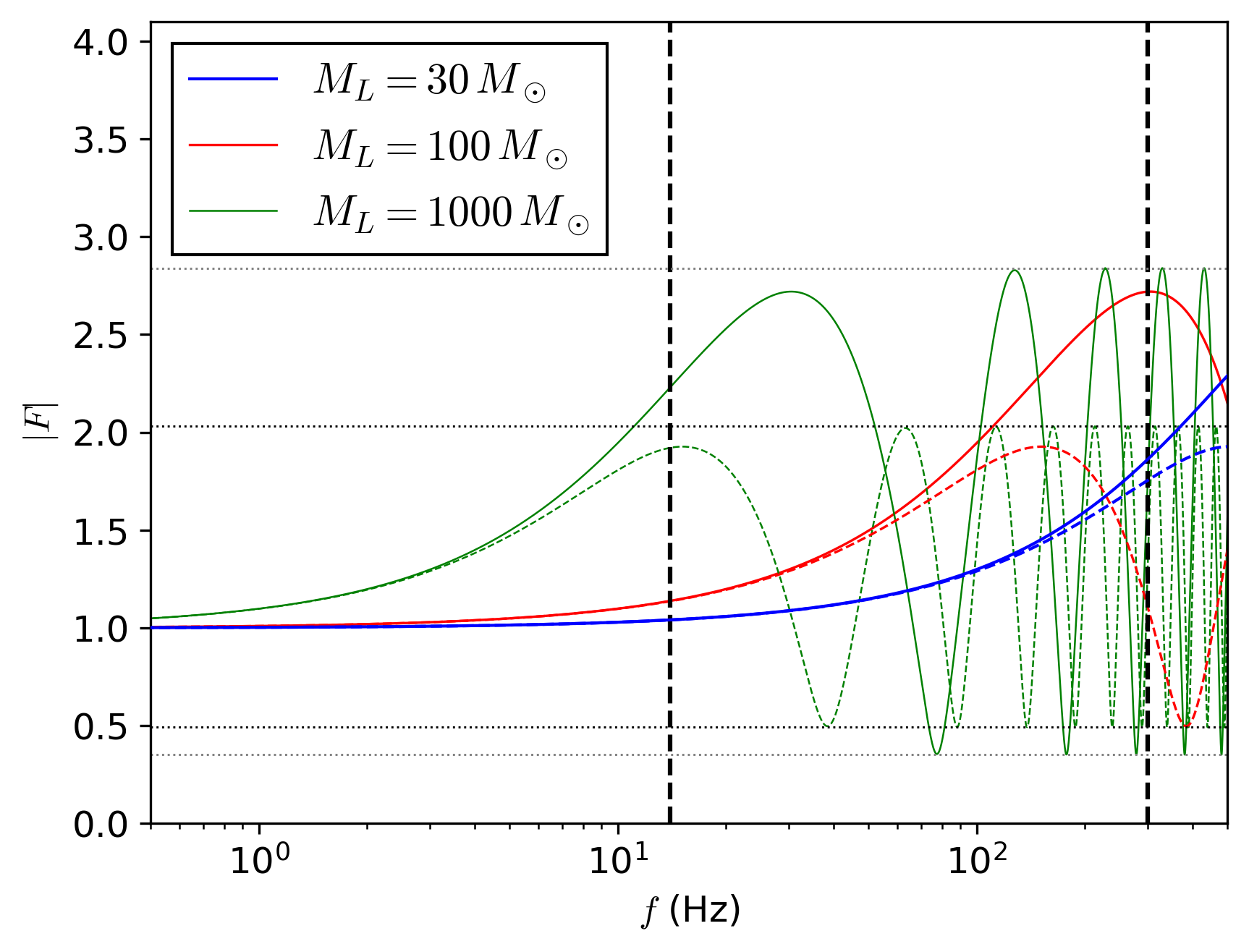}
\caption{Transmission factor as a function of frequency for different values of $M_L$ at two fixed values of $y$: $y=0.25$ (solid curves) and $y=0.5$ (dashed curves).  The black vertical dashed lines correspond to $f= 15 Hz$, given by the sensitivity of the detector, and $f=300$ Hz, the merger frequency for a $M_{det} = M_1 + M_2 = 60$ $M_\odot$ BH binary.  
}
\label{fig:regions}
\end{figure}

For calculations that are numerically expensive we use a hybrid transmission factor function that takes the value of the full-wave $F$ from Eq.~\eqref{F_PML} at low frequencies
and its GO limit given by Eq.~\eqref{F_GO_amp1} at high frequencies.
The simplicity of the GO formula makes the computation much faster. 
As a matching point between these solutions, we take the frequency of the third oscillation maximum to ensure a smooth transition.
\subsection{SNR increase}
\label{SNR-section}
The detectability of a GW signal may be characterized by the signal-to-noise ratio (SNR).
We first define a noise-weighted inner product between frequency-domain waveforms \cite{cutler94,flanagan98}
\begin{equation}
\langle a, b \rangle = 2 \int_{f_{\rm min}}^ {f_{\rm max}}
\frac{\widetilde{a}^{*}(f) \widetilde{b}(f) + \widetilde{a}(f) \widetilde{b}^{*}(f)}{S_n(f)} df ~,
\end{equation}
where $S_n(f)$ in the one-sided noise power spectral density, and the minimum and maximum frequencies of the detector $f_{\rm min}$ and $f_{\rm max}$ are chosen to contain all the practically detectable power in the waveform.

We also define moments of the noise spectral density in a manner similar to \cite{PoissonWill95}
\begin{equation}
K_\alpha = \left( \int \frac{f^\alpha  \, |h(f)|^2}{S_n(f)} ~ \mathrm{d}f \right) ~\Bigg/~ \left( \int \frac{|h(f)|^2}{S_n(f)} ~ \mathrm{d}f \right) ~.
\label{eq:Ka}
\end{equation}
These moments characterize the sensitivity of a given detector to changes in the waveform. For instance, the sensitivity to time offsets is governed by $K_1$ and $K_2$, because a time derivative is equivalent to multiplication by $f$ in the frequency domain. We will use these moments in predicting the SNR and match in the amplification region.

The SNR is then defined by
\begin{equation}
\rho^2 = \langle h, h \rangle = 4 \int_{f_{\rm min}}^ {f_{\rm max}} \frac{|\widetilde{h}(f)|^2}{S_n(f)} df ~.
\end{equation}

The relative SNR increase due to lensing can then be written as the ratio between the lensed and unlensed SNRs:
\begin{equation}
    \rho_{\rm rel}^2 = \frac{\langle h, h \rangle}{\langle h_{UL}, h_{UL} \rangle}.
\label{eq:snr_increase}
\end{equation}
We will primarily consider gravitational waves from binary black hole coalescence since they are the most massive and the furthest away from the detector, and thus the most likely to be lensed.  

\begin{figure}[!htbp]
\includegraphics[width=1.0\columnwidth]{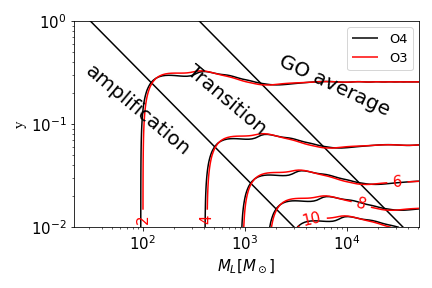}
\caption{Relative increase in signal-to-noise ratio $\rho_{\rm rel}$ as a function of $y$ and $M_L$ for the same black hole binary of $M_{det} =M_1 + M_2 = 60$ $M_\odot$ for the O3 and O4 LIGO power spectra densities. The oblique black lines delimit the regions where the amplification and the GO approximations are valid, and the transition between them. 
}
\label{fig:snr-transition}
\end{figure}

\begin{figure}[!htbp]
\includegraphics[width=1.0\columnwidth]{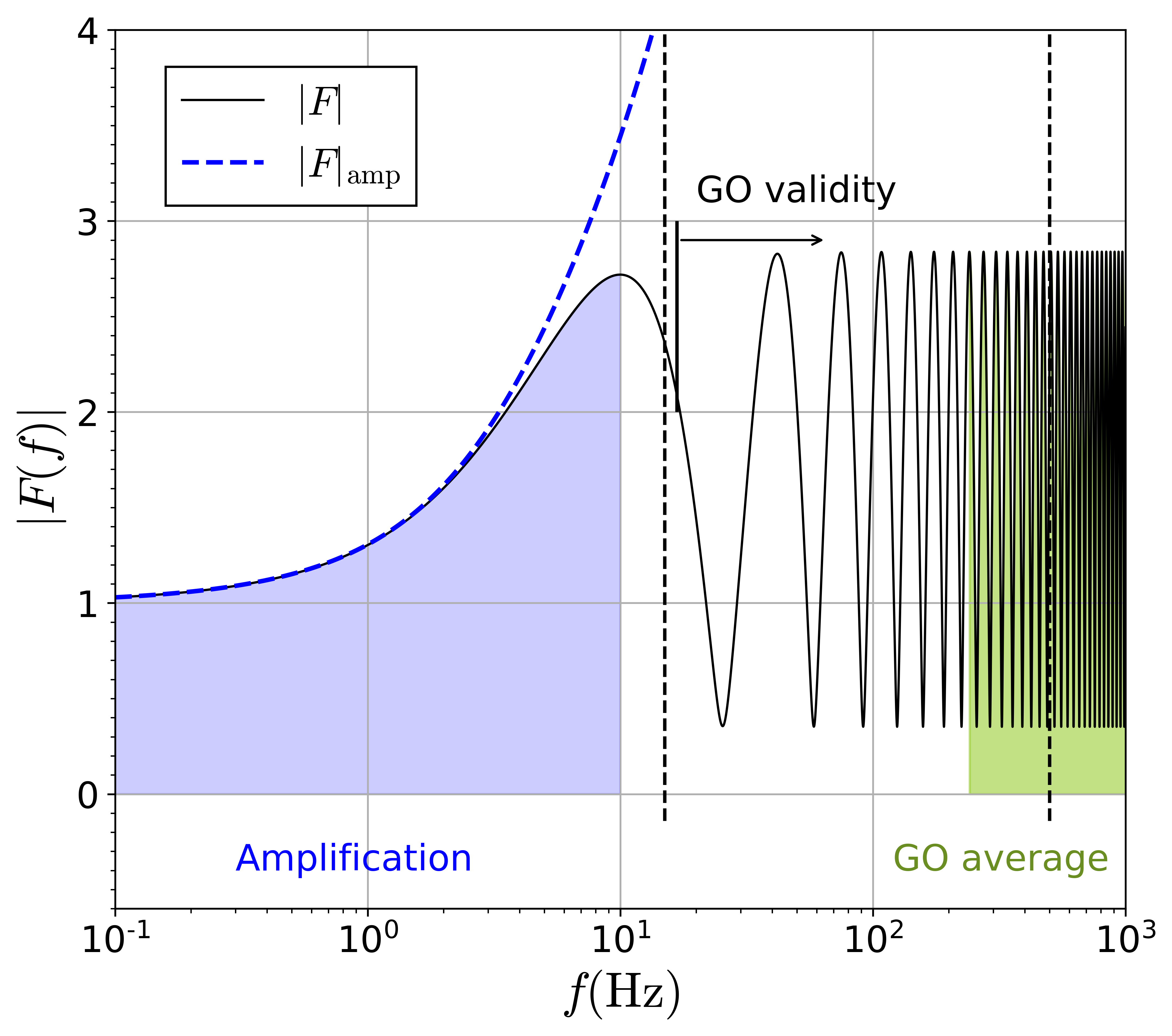}
\caption{Transmission factor $|F|$ as a function of frequency $f$, for fixed values of $M_L = 3000 M_\odot$ and $y=0.25$. The LVK detectability lies between the black dashed lines with a maximum $f = 500$ Hz for illustrative purposes. 
There are two colored regions of interest related to the SNR increase $\rho_{\rm rel}$} shown in Fig.~\ref{fig:snr-transition}: (i) the amplification region [Eq.~\eqref{abs_Famp_exact}] and (ii) the GO average region [Eq.~\eqref{eq:rho_asympt}]. 

\label{fig:3regions}
\end{figure}

We consider gravitational wave strain amplitudes obtained with LVK's IMRPhenomD waveform \citep{Husa-16, Khan-16} that go through the standard inspiral, merger and ringdown phases. 
The effect of lensing on the SNR increase is demonstrated in Fig.~\ref{fig:snr-transition}.
We take the usual $M_{\rm det} = M_1 + M_2 = 60 M_\odot$ binary and represent the results in terms of two lensing parameters, $M_L$ and $y$.
Since it is a ratio, $\rho_{\rm rel}$ is roughly independent of the LVK noise curve used showing very similar results for O3 and O4\cite{O3O4}.
 The characteristic behavior in different regions can be understood by comparing with the transmission factor represented in Fig.~\ref{fig:3regions}. The boundary between the amplification and transition region is chosen when the amplification approximations starts to diverge at $f t_M y \approx 0.055$ \cite{BU-22}, and the boundary with the GO average region is taken to be at the 8th maximum. However, these can be varied depending on the accuracy needed.
  

\section{Detected SNR}
{   While in this paper we are most interested in first detections, which are likely to happen in O4 or O5 for close-to-threshold events with $\rho = 8-10$, for sufficiently high SNR systems even small mismatches would lead to observable effects. 

In searches, gravitational wave signals are penalized for mismatching. Pipelines use statistics to downweight signals that do not look like templates in their bank to favor real candidates over transient non-Gaussian noise sources known as glitches.

The detected SNR is reweighted when $\chi_r^2 > 1$ to
\begin{equation}
    \hat{\rho} = \rho/[1 + (\chi_r^2)^3/2]^{1/6},
\end{equation}
where $\chi_r^2 = \chi/(2 p -2)^2$, $\chi \leq \chi_{\rm max} = 2(p-1) + 2 \rho^2 (1-{\cal M})$, ${\cal M}$ is the match and $p$ is the number of bins \cite{Usman-15,Messick-17,allen-05}. 

The mismatch makes the search less sensitive by decreasing the detected SNR and by increasing the $\chi^2$. In our case, for a mismatch of $1-{\cal M} = 10$\% at $M_{\rm det} = 60 M_\odot$, a gravitational wave signal with $\rho = 10$ will be detected as $\hat{\rho} \approx 8.5$. The SNR will be reduced further for higher $M_{\rm det}$ since the detectable part of the waveform shortens, which reduces the number of frequency bins and increases the penalty on the waveform.

 The $\chi^2$-statistics  is performed as part of the lensing search, and is beyond the scope of this work where we only provide upper limits for lensing detection.
 }

\section{Waveform mismatch: matching lensed events to existent templates}
\label{match}
\begin{figure*}[!htbp]
\includegraphics[width=1.0\columnwidth]{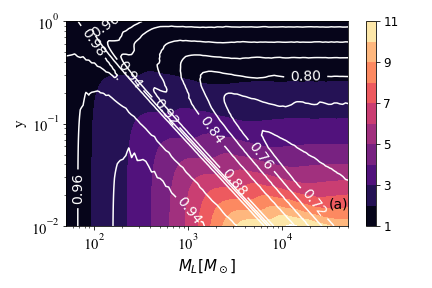}
\includegraphics[width=1.0\columnwidth]{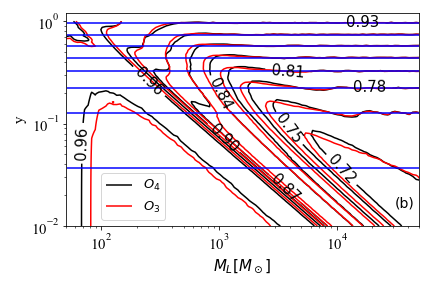}
\caption{Match between a lensed and unlensed waveform from a $M_{det} = 60 M_\odot$ black hole binary (a) superimposed on the SNR increase computed for O4 (b) for O3 and O4 power spectra densities. It can be seen that the highest mismatch occurs for high SNR increase. The solid blue lines are the match computed in the GO limit from Eq.~(\ref{eq:MatchGO}). They are equally spaced contours between 0.72 (lowest blue line) and 0.93 (highest blue line). The agreement is poor for low $y$ and $M_L$. Moreover, the change from O4 to O3 has minor effects.} 
\label{fig:snr-match}
\end{figure*} 

\begin{figure*} [!htbp]
\includegraphics[width=1.0\columnwidth]{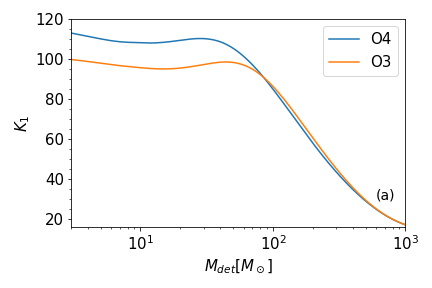}
\includegraphics[width=1.0\columnwidth]{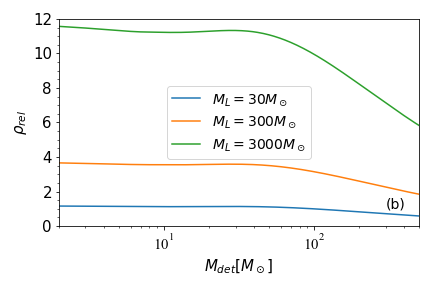}
\includegraphics[width=1.0\columnwidth]{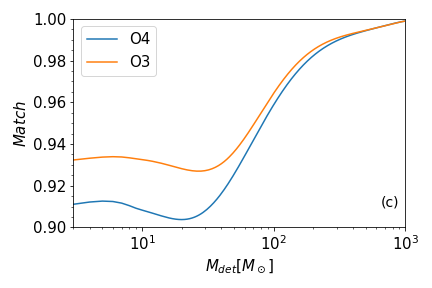}
\caption{(a) the first SNR moment $K_1$, (b) $\rho_{rel}$ and (c) match in the amplification region for O3 and O4 power spectra densities as a function of total mass $M_{det}= M_1 + M_2$.} 
\label{fig:momRel}
\end{figure*} 

Microlensing is not just an overall amplification of the
signal, but a frequency-dependent change to the amplitude and phase of the gravitational wave. This intrinsic distortion of the signal
is what will provide our evidence of lensing. The degree
of similarity between two waveforms is quantified by the
match, which ranges from zero to one. The match is the overlap defined below maximized over phase and time, as this is what the detection process will do.
The detectability of a given lens is computed by finding the optimal match between the lensed waveform and unlensed waveforms in the LVK's template bank. The SNR of a detection made with the unlensed template bank will be reduced proportionally to the match. In what follows, we will also use the mismatch, defined as one minus the match.

The overlap between two waveforms $a$ and $b$ is defined as
\begin{equation}
    {\cal O}(a, b) = \frac{\langle a, b \rangle}{\sqrt{\langle a, a \rangle \, \langle b, b \rangle}} ~.
\end{equation}
Phase and time are incorporated into a waveform as $h(f; t, \phi) = h(f) e^{-i (2 \pi f t + \phi)}$. The match is the overlap maximized over time and phase offsets between two waveforms
\begin{eqnarray}
  \label{eq:match_definition}
    {\cal M}(a, b) &=&  \max \limits_{\phi,t} \frac{\langle a(f) , b(f; t, \phi) \rangle}{\sqrt{\langle a, a \rangle \, \langle b, b \rangle}} ~.
 \end{eqnarray}
The match is the appropriate quantity for comparing the similarity of two waveforms, as the detection process will always search over (maximize) the unknown time and phase.
 
 In the amplification region, 
we can use Eq.\ (\ref{rhoamp}) to simplify the match to
 \begin{eqnarray}
 \label{eq:OverlapAmp}
 {\cal M}(h,h_{UL}) &=&  
 \frac{\langle f^{1/2} h_{UL},  h_{UL} \rangle}{\sqrt{\langle h_{UL}, h_{UL} \rangle \, \langle \sqrt{f} h_{UL},  \sqrt{f} h_{UL} \rangle }} \\ \nonumber &=& \frac{K_{1/2}}{\sqrt{K_1} },
 \end{eqnarray}
 where $K_{1/2}$ and $K_1$ are defined in Eq.\ (\ref{eq:Ka}). The integral is maximized when the time and phase offsets are zero because the integrand is real-valued. The factors of $t_M$ in the lensed waveform cancel, thus in this regime the overlap is independent of both $y$ and $M_L$. Fig.\ \ref{fig:momRel}(c) displays the match as a function of $M_{det}$. For $M_{det}=60 M_\odot$, the match is $0.943$ for O3 and $0.934$ for O4; the waveforms are more distinguishable since the noise curve is wider and flatter. This approximation agrees with our numerical calculations for the match in the amplification region. 
 
 In the GO average region, two images $h_{GO} = h_1 + h_2$ begin to appear, separated by a time delay. Because of the time delay, the unlensed template can only align with one of the two images, so the match is determined by the magnitude of the first (stronger) image relative to the sum of the magnitude of both images
\begin{eqnarray}
  \label{eq:MatchGO}
    {\cal M}(h_{GO}, h_{UL}) &=&  \max \limits_{\phi,t} \, \frac{\langle h_1 ,h_{UL} \rangle}{\sqrt{\langle h_{UL} ,h_{UL} \rangle ~ \langle h_{GO} , h_{GO} \rangle}}  \\ \nonumber
&& = \frac{\sqrt{\mu_1}}{\sqrt{\mu_1+ \mu_2}},
 \end{eqnarray}
with the magnification of each of the images being
\begin{equation}
\mu_{1,2}=\displaystyle{\frac{1}{4} 
\left( \frac{y}{\sqrt{y^2+4}} + \frac{\sqrt{y^2+4}}{y} \pm 2 \right) }.
\label{eq:magnifications}
\end{equation}
We have used the fact that $\langle h_1, h_2 \rangle \approx 0$ because the two images are not aligned in time, and $h_{1,2} \propto \sqrt{\mu_{1,2}} h_{UL}$ plus the time shift (and phase shift due to the Morse factor). 

Fig.~\ref{fig:snr-match} shows  the match between an unlensed and a lensed equal mass $M_{det} = 60 M_\odot$ black hole binary. We use the LVK's standard IMRPhenomD templates with no spin, precession or eccentricity included. Preliminary results that included optimization over mass and spin showed a difference of under 1~\%. We so far find that the optimal match occurs between the lensed $h$ and unlensed waveform $h_{UL}$ of the same binary mass. However, a more exhaustive parameter study is needed. In Fig.~\ref{fig:snr-match}(a) we place the increase in SNR and the match on the same plot in $y-M_L$ space. The two regions: amplification (left corner) and GO (right corner) can clearly be distinguished. 

If the source has $\rho\approx 10$, a mismatch higher than 20\% can bring it just under the detectability threshold $\rho=8$, rendering it undetectable. As we will see in the next section, sources with lower 'actual' $\rho$ are more likely to be lensed because they are expected to be further away. As per the figure, a distant source of $\rho_{UL}=1$ could increase its SNR by a factor of $10$ and appear detectable at $\rho=8$ with a match of 84\%.

Fig.~\ref{fig:snr-match}(b) compares the optimal match computed for the same $M_{det} = 60 M_\odot$ binary for O3 and predicted O4 noise curves. As shown mathematically, both the match and SNR increase are largely insensitive to the noise curve used. In the GO region, they only depend on $y$. In the silent amplification region, where the match is generally over 90\%, some difference is observed due to variations in the first SNR moment $K_1$ (see Fig.~\ref{fig:momRel}(a) and (b) for $K1/\rho_{UL}$ and $\rho_{rel}$). Minor variations in the match are also observed in the region where the full transmission factor is required for the calculation. As $M_L$ increases, the oblique lines become denser and the binary moves from the amplification into the GO region. The  horizontal blue lines show when the analytic formula given by Eq.~(\ref{eq:MatchGO}) from the GO estimate matches the full expression. Since we are plotting a ratio, there is no need to choose $\rho_{UL}$ for this figure.


\section{Probability of detection: An Upper Limit}
\label{rates}
{  In this section, we derive the relative probability of detecting a lensed vs an unlensed source assuming a constant number of sources per comoving volume. Detection here means that the SNR is above a set threshold. This gives us the relative fraction of events that will be lensed. We will relate this to the ratio of prior probabilities that is used to convert the Bayes factor into a posterior probability ratio.}

The distribution of lenses in angle is scale-free for lenses sufficiently near the line of sight, i.e., independent of $M_L$. The lensing angles are distributed uniformly in area, as $y dy$. {   But there is a strong selection bias for smaller values of $y$ because they lead to substantial magnification. The magnification increases the volume containing sources, as well as  the probability of encountering a lens due to the greater distance. This leads to the the distribution of {\it detected} lens parameters to strongly favor smaller $y$. This effect must be considered when simulating lensing and interpreting the results.} 
Our derivation will follow the lines of \cite{TurnerOstriker1984} (and see \cite{urrutia22} for an application to gravitational waves), but adding the effect of selection bias due to detectability above an SNR threshold. 

The probability of lensing depends on the relative proportion of lensed to unlensed detections,
\begin{equation}
F_{\mathrm{lensing}}(M_L,n_L) = \frac{N_{LS}}{N_S},
\label{eq:proportion_lensing}
\end{equation}
where $N_{LS}$ is the number of lensed sources above the threshold SNR $\rho_t$, and likewise $N_S$ is the number of unlensed detectable sources. We will consider below a few different criteria for a source to be meaningfully lensed. We treat the GW sources as uniformly distributed in comoving volume with number density $n_S$. We observe that the  relative probability of detection is independent of any overall scaling of $n_S$. $F_{\mathrm{lensing}}$ scales directly with the lens density $n_L$. The probability that a given detection is lensed is
$P_{\mathrm{lensing}} = F_{\mathrm{lensing}}(M_L,n_L) / (1 + F_{\mathrm{lensing}}(M_L,n_L))$,
which is almost equal to $F_{\mathrm{lensing}}(M_L,n_L)$ for $N_{LS} \ll N_S$.

To incorporate cosmology, we will write our integrals in terms of the radial comoving distance
\begin{equation}
\chi = \int \frac{dz'}{H(z')}
\end{equation}
where $H(z)$ is the Hubble parameter. We assume standard cosmology
\begin{equation}
H(z) = H_0 \sqrt{\Omega_M (1+z)^3 + \Omega_\Lambda},
\end{equation}
with $H_0 = 69.6$ km s$^{-1}$ Mpc$^{-1}$, $\Omega_M = 0.286$, $\Omega_\Lambda = 0.714$. 
 As the universe is spatially flat, the transverse and radial comoving distances are identical. The volume integrand is simply $dV = \chi^2 d\chi d\Omega$.
  In this section, we use $M_L$ to mean the intrinsic mass of the lens, and the redshift into the observer frame is written explicitly as $(1 + z_L) M_L$.

The total number of sources with SNR above a detection threshold $\rho_t$ can be written as
\begin{equation}
    N_S = 4 \pi \int_0^{\chi_t} n_S \,\chi_S^2 \, 
    d\chi_S = \frac{4 \pi}{3} \chi_t^3 \,n_S.
\label{eq:unlensed_number_integral}
\end{equation}
Above, $\chi_t$ is the comoving distance at which the unlensed source would have SNR above a threshold, which in this manuscript is $\rho_t = 10$. For simplicity, we are assuming that the comoving number density of lenses $n_S$ is independent of redshift, and also of source mass when we are considering detections where the observed mass (in the detector frame) is fixed. These effects can be included, though we note that the scenarios we consider do not extend past $z_S \approx 4$, where the density would be expected to drop. In general, constant comoving density for both sources and lenses is appropriate for e.g., primordial black holes in the absence of mergers. 

We next compute the number of lensed objects $N_{LS}$ above a threshold $\rho_t$ as a function of source redshift $z_S$ and lens parameters $y$ and $M_L$. This involves an integral over both the lense and source location

\begin{equation}
N_{LS} = \int n_S \int n_L dV_L dV_S ~.
\end{equation}

Here $n_L$ is the number density of lenses in a comoving volume, and we have
\begin{equation}
n_L ~ dV_L = 2 \pi n_L \chi_L^2 \sin \theta_L d\theta_L d\chi_L. 
\end{equation}
The angle $\theta_L$ will be very small, so that $\sin \theta_L \approx \theta_L$.
This is related to the Einstein angle by $\theta_L = \theta_E \, y$ (see Sec.\ \ref{sec_pml}).


The angular diameter distances can be expressed in terms of $\chi$ as $d_S = \chi_S / (1+z_S)$ and $d_L = \chi_L / (1+z_L)$. The distance from lens to source is \cite{urrutia22}
\begin{equation}
d_{LS} = d_S - \frac{1+z_L}{1+z_S} d_L = \frac{\chi_S - \chi_L}{1+z_S}.
\end{equation}.

We now change variables from $\theta_L$ to $y$ to obtain the number of lenses closer than the source
\begin{eqnarray}
N_\mathrm{L} &=& 8 \pi \int n_L ~ M_L \frac{d_{LS}}{d_L d_S} y \, dy \, \chi_L^2 \, d\chi_L \\
&=& 8 \pi \int n_L ~ M_L \frac{(1+z_L) (\chi_S - \chi_L) \chi_L}{\chi_S} 
y \, dy \, d\chi_L .
\end{eqnarray}
The integrand can be seen as a differential form of the standard lensing cross section.

Integrating over all detectable sources yields the total number of lensed sources
\begin{eqnarray}
N_\mathrm{LS} &=& 4 \pi \int n_S N_L \chi_S^2 \, d\chi_S \\
&=& \int_0^{y_{\rm max}} dy \int_0^{\chi_t(y)} d\chi_S \int_0^{\chi_S} d\chi_L \, p(y, \chi_L, \chi_S)
\label{eq:total_lensed_sources}
\end{eqnarray}
where
\begin{equation}
p(y, \chi_L, \chi_S) = 32 \pi^2 n_S n_L \, M_L \, (1+z_L) \, \chi_L (\chi_S - \chi_L) \,\chi_S \, y \,
\label{eq:probability_density}
\end{equation}
and $N_L$ is the number of lenses.

\begin{figure*}[!htbp]
\includegraphics[width=2.0\columnwidth]{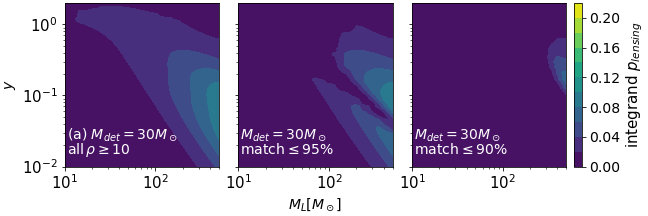}
\includegraphics[width=2.0\columnwidth]{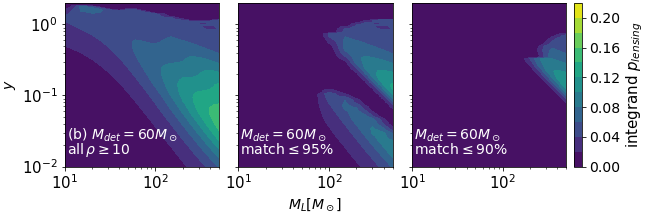}
\includegraphics[width=2.0\columnwidth]{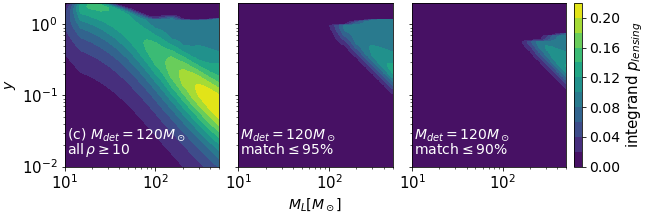}
\caption{Probability density as a function of $y$ and $M_L$ for black hole binaries of total detected mass (a) $M_{det}= 30 M_\odot$, (b)  $M_{det}= 60 M_\odot$ and (c)  $M_{det}= 120 M_\odot$ for all sources of $\rho \ge 10$. The first column plots all lensed sources, the second the lensed sources with a mismatch of 5\% or greater, and the third the lensed sources with a mismatch of 10\% or greater. All are scaled by the total number of sources. The GO average and amplification only regions separate for intermediate mismatch $\geq 5$ \%. It can be seen that the most lensed sources occur for $M_{det}= 120 M_\odot$. While many sources may be lensed only the fraction of them that have detectable mismatch with unlensed templates can be identified as lensed by LVK searches.} 
\label{fig:Pdensity}
\end{figure*} 

\begin{figure*}[!htbp]
\includegraphics[width=2.0\columnwidth]{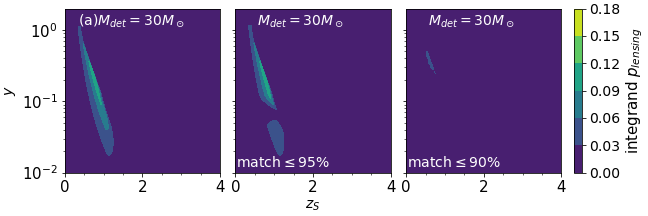}
\includegraphics[width=2.0\columnwidth]{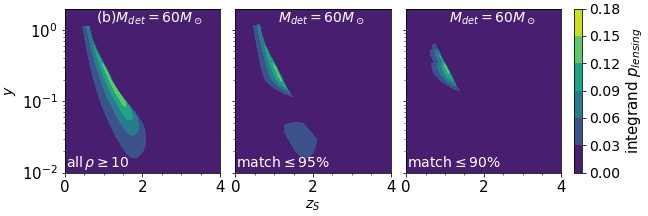}
\includegraphics[width=2.0\columnwidth]{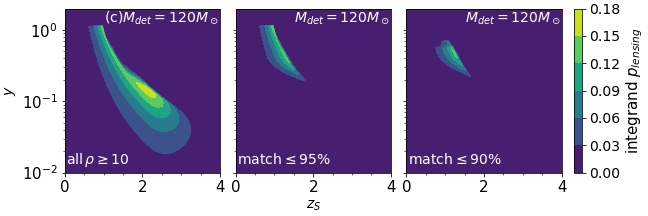}
\caption{Probability density of $M_L= 300 M_\odot$ as a function of $y$ and $z_S$ for black hole binaries of total detected mass  (a)$M_{det}= 30 M_\odot$, (b)$M_{det}= 60 M_\odot$ and (c)  $M_{det}= 120 M_\odot$ for all sources of $\rho \ge 10$. The first column shows all lensed sources, the second the lensed sources with a mismatch of 5\% or greater, and the third the lensed sources with a mismatch of 10\% or greater. All are scaled by the total number of sources. Like before it can be seen that the most lensed sources occur for $M_{det}= 120 M_\odot$ which can be seen up to $z \approx 3.2$, and that only a small number of sources have a mismatch of 5\% or more ($z_S<2$). When the mismatch increases to 10\% or more, the amplification only region disappears.} 
\label{fig:Integrand}
\end{figure*} 

\begin{figure}[!htbp]
\begin{center}
\includegraphics[width=1.0\columnwidth]{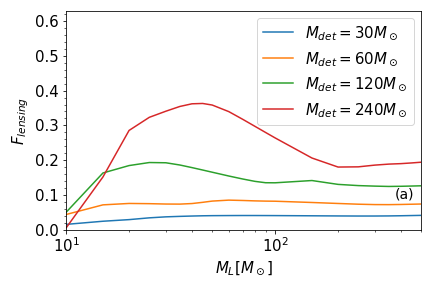}
\includegraphics[width=1.0\columnwidth]{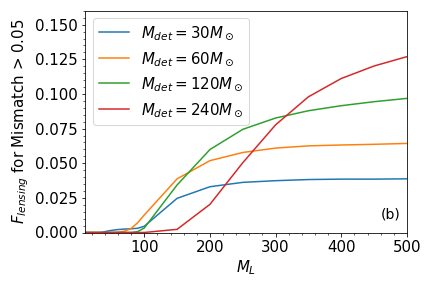}
\includegraphics[width=1.0\columnwidth]{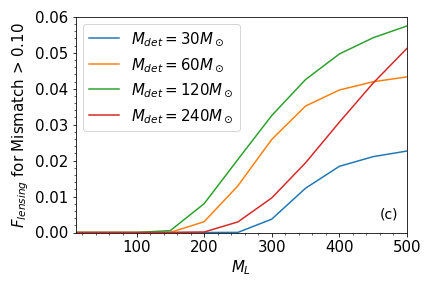}
\end{center}
\caption{Proportion of lensed detections $F_{\mathrm{lensing}}$ (Eq. \ref{eq:proportion_lensing}) shown as a function of $M_L$ for (a) all lensed sources, (b) lensed sources with a mismatch $\ge$ 5\%  (c) lensed sources with a mismatch $\ge$ 10\% for black hole binaries of total detected mass $M_{det}= 30 M_\odot$, $M_{det}= 60 M_\odot$, $M_{det}= 120 M_\odot$ and $M_{det}= 240 M_\odot$ computed for $\rho \ge 10$. In these plots, the number of lenses is scaled so that they comprise all of the dark matter density of the universe. It can be seen lensing becomes discernible with the LVK only for $M_L \gtrsim 300 M_\odot$ with less than 10\% of lensed events identifiable as lensed.} 
\label{fig:Probability}
\end{figure}

The constraint $\chi_L<\chi_S$ appears because the lens must not be beyond the source.

Importantly, the amount of lensing changes the distance to which we can observe a source.  The relative SNR increase is a function of $y$ and the redshifted lens mass $(1+z_L) M_L$. In addition to the value of this integral, we will also examine the integrand of Eq.\ (\ref{eq:total_lensed_sources}) as a function of $y$ to determine the distribution of $y$ among systems with detectable lensing. Regions of parameter space with more lensing amplification will give larger contributions to the integral. 
The integrand can be used to determine the distribution of detectable sources as a function of $y$.

We perform our integrals by grid integration over $\chi_S$, $\chi_L$, and $\log y$; a Monte Carlo approach would be required if we considered more variables. We take $y_{\rm max} = 2$, which goes beyond the point where the lensing contribution is significant. 

We consider the density of lenses to be the same as that of the dark matter in the universe $n_L M_L = \rho^{DM}_{\rm cr}$, which provides a strict upper limit for the lenses of a given mass. Since we cannot have more lenses than dark matter, this means that the probability that a certain detected black hole binary system is lensed can be lower than the numbers in this manuscript, but it cannot be higher unless the model changes e.g., the point mass lens is embedded into a larger object that can increase the lensing. The probability is independent of $n_S$ unless it evolves significantly in the relevant redshift range.




The gravitational waves the LVK measures are at the location of the detector. 
In the detector frame,
\begin{equation}
h \propto \frac{(\mathcal{M}_{det})^{5/6}}{D_{\rm lum}}
= \frac{(\mathcal{M}_{det})^{5/6}}{(1+z_S) \chi_S},
\end{equation}
where $\mathcal{M}_{det}$ is the measured (detector frame) Chirp mass of the binary. The source-frame (intrinsic) Chirp mass is related by $\mathcal{M}_{det} = (1+z_S) \mathcal{M}_{\rm src}$.


We then visualize the distribution of lensed sources as a function of $y$, which is Eq.\ (\ref{eq:probability_density}) marginalized over the lens and source positions:
\begin{equation}
p(y) = \frac{1}{N_{LS}} \int_0^{\chi_t(y)} d\chi_S \int_0^{\chi_S} d\chi_L \, p(y, \chi_L, \chi_S) ~,
\end{equation}
 
The lenses themselves are distributed like $p(y) \propto y$, but the prior is modified by some lensing configurations being \emph{more detectable} than others. This causes the {   distribution to peak well below $y = 1$, and is further restricted if we require the lensing to be detectable}, i.e., to have enough mismatch with the unlensed waveform.{   We note that small mismatch can be significant at high SNR. } 

Fig.\ \ref{fig:Pdensity} displays the probability density $p(y)$ as a function of lens mass $M_L$ normalized by the density of unlensed sources $N_S$, integrated over the lens position $\chi_L$. As expected, more massive binaries have a higher probability of being lensed. Figs. \ref{fig:Pdensity}(a)-(c) show the probability density for black hole binary systems of total detected mass $M_{det} = M_1 + M_2 = 30 M_\odot$, $M_{det} = 60 M_\odot$ and $M_{det} = 120 M_\odot$. The second column of each figure shows the same probability density restricted to when the mismatch is 5\% or higher, while the third column restricts the mismatch to 10\% or higher. The higher the mismatch with an unlensed source, the more likely it is to identify a source as being lensed. Otherwise, lensing is silent. In the second column of Fig.\ \ref{fig:Pdensity}(a) and (b), the probability density clearly separates in two regions. The higher $y$ region occurs when Geometrical Optics is valid, and the lower $y$ region corresponds to the amplification only scenario.  As the total mass of the binary increases to $M_{det} = 120 M_\odot$ (Fig. \ref{fig:Pdensity}(c)), only the GO region is discernible reducing the $y-M_L$ range of identifiable lensed systems to a small area in the upper right corner of the plane.

Fig. \ref{fig:Integrand} plots the probability density from Eq.\ (\ref{eq:probability_density}) for a given lens mass $M_L = 300 M_\odot$ as a function of $y$ and source redshift $z_S$, {   integrated over} the lens position $\chi_L$. Like before, Figs. \ref{fig:Integrand}(b) and (c) (second plot on each line) shows the two separate regions for a mismatch of 5\% or higher. Higher $M_{det}$ reaches higher source redshift. For $M_{det} = 120 M_\odot$ (Fig. \ref{fig:Integrand}(c)), potential lensed sources go up to $z_S \approx 3.2$. Unfortunately, the higher redshift systems have low mismatch and cannot be identified as lensed. For the standard $M_{det} = 60 M_\odot$ (Fig. \ref{fig:Integrand}(b)), the highest $z_S \approx 2$ corresponds to the amplification only region.  At this redshift  $M_{det} = 60 M_\odot$ results in a total mass of $M_{src} = 20 M_\odot$ or $M_1 = M_2 = 10 M_\odot$ in the source frame, which is closer in mass to the black holes observed in our galaxy with electromagnetic observations. The probability of lensing for such an event would be around $3\%$ if all dark matter was composed of black holes of $M_L = 300 M_\odot$. 

After the integrals over $y$, $\chi_S$ and $\chi_L$ are performed, Fig.~\ref{fig:Probability} shows the probability of lensing as a function of $M_L$ for $M_{det} = 30 M_\odot$ to $M_{det} = 240 M_\odot$ with $\rho \ge 10$. Like before, it is normalized by the total number of sources with an $\rho  \ge 10$. Fig.~\ref{fig:Probability}(a) shows the probability for all lensed sources. We can see it peaks at low lens mass, and that the peak shifts to slightly higher masses as $M_{det}$ increases reaching $M_L \approx 40 M\odot$ for $M_{det} = 240 M_\odot$, and then flattens out as the lens mass continues to increase. In Figs.~\ref{fig:Probability}(b) and (c) we limit the detected black hole binaries to those whose waveform has a mismatch of $\ge 5 \%$ and $\ge 10\%$, respectively. This decreases the probability of lensing to a value that is well below its peak. It can be seen that detectable lensing starts at larger values of $M_L >100 M_\odot$ and flatten out as $M_L$ increases. For a detectable mismatch of $\ge 10\%$, increasing the detected mass from $M_{det} = 30 M_\odot$ to $M_{det} = 60 M_\odot$ increases the probability of detectable lensing by about a factor of $2$. However, as the detectable mass increases further, the increase in probability is less significant. Already for $M_{det} = 240 M_\odot$ contains lower probability events than  $M_{det} = 120 M_\odot$.

{  
We now consider the Bayesian comparison of the hypothesis that an observed signal is lensed versus unlensed. For two hypotheses $\mathcal{H}_1$ and $\mathcal{H}_2$, the ratio of posterior probabilities is

\begin{equation}
\frac{p(\mathcal{H}_2|d)}{p(\mathcal{H}_1|d)} = \frac{p(\mathcal{H}_2)}{p(\mathcal{H}_1)} \times
\frac{p(d|\mathcal{H}_2)}{p(d|\mathcal{H}_1)}
\end{equation}

The second term, the Bayes factor, is commonly computed by nested sampling which yields the evidences

\begin{equation}
p(d|\mathcal{H}_i) = \int p(d|\theta_i, \mathcal{H}_i) \; p(\theta|\mathcal{H}_i) \mathrm{d} \; \theta_i ~.
\end{equation}

The first term is the prior probability ratio of the two hypotheses as derived above. This is required for meaningful interpretation of the Bayes factor. In our case, we need to assume a certain density of lenses to find the prior probability, and this changes the interpretation of the Bayes factor. For a fixed Bayes factor, the posterior probability of lensing increases proportional to the assumed density of lenses.

The method typically used \cite{hannuksela19,lensingO3a} is for the prior used in the evidence calculation to be over a fixed volume, rather than conditioned on detectable systems. Then the ratio of prior probabilities should make the same choice. The result will be nearly the same as with the conditioned prior because the evidence only accumulates contributions where the likelihood is high\footnote{We thank an anonymous referee for pointing this out.}. 

We note some important points when computing the evidences. First, as seen in Fig.~\ref{fig:Integrand}, lenses can be detected to a much larger distance than unlensed systems, and the distance prior used must support this. Next is the prior on $y$. We see in Fig.~\ref{fig:Integrand} that lensing detections are common at small $y$ and so the prior must not be artificially cut off there.

On the other hand, at large $y$ the waveform becomes indistinguishable from the unlensed one. In this case, the likelihood of lensing $\Lambda_\ell(y)$ is effectively the same as the unlensed likelihood $\Lambda_{ul}$. If this occurs at a value $y_*$, the evidence integral breaks into a piece below $y_*$ and a piece above. Suppressing the other variables the evidence is

\begin{equation}
    E_\ell = \left( \int_{y_{min}}^{y_*} \Lambda_\ell(y) y \, \mathrm{d}y + \int_{y_*}^{y_{max}} \Lambda_{ul} y \, \mathrm{d}y \right) \Big/ \int_{y_{min}}^{y_{max}} y \, \mathrm{d}y
\end{equation}

Because of the normalization, the values of $y$ which are effectively unlensed dilute the evidence with the second term, which is the same as the unlensed evidence. Hence restricting the focus only to significantly lensed systems makes this test more sensitive.
}

{   In this manuscript we prioritized first detections from O4 and O5, which are likely to be events close to the detectability threshold $\rho = 8-10$.  For these modest SNRs, 
mismatches of $\gtrsim 5\%$ are  likely necessary to support the detection of lenses. However, for sufficiently high $SNR$ even small mismatches can lead to observable effects. This is different from the case
of other subtle effects like precession, because it is possible to have a closer and louder precessing event where we could  discern it, but lensing is less likely for closer events. }

{   Other works \cite{jung19,basak22, urrutia22} that study  lensing signatures to investigate if primordial black holes make up a fraction of all the dark matter
in the Universe
use
a cross-section for lensing that is of the order of the Einstein
radius. They compute the rates of being within this impact parameter as a
function of source redshift, and integrate out to a horizon set by the SNR threshold for un-lensed events (in the case of Basak et. al. \cite{basak22}, this
is a feature of the simulation they lay out in their Appendix).
However,  when considering the rates of detected events, which we take to be ${\rm SNR} > 10$, they do not take into account that horizon is ``pushed out" as
a function of impact parameter $y$. This effect is taken into account in our probability distribution. 

A complete search that considers a network of detectors as well as the inclination of the source and the response of each detector is beyond the purpose of this first paper. The effects from the source orientation and response of the detector can be significant. We do not take into sky position or the inclination of the source \cite{ng18}, which change the response of the detector. This will be done in subsequent work.}

\section{Conclusions and Future Directions}
\label{conclusions}
To date, the LVK collaboration has detected over 90 compact binary systems. The detectors' sensitivity can be expected to improve over the coming years. So far it appears that the black hole population seen by the LVK is an order of magnitude heavier than black holes found by X-ray surveys of the Milky Way. It is known that a population could appear more massive and less distant if redshift is underestimated. Such a binary black hole population could then provide first indirect evidence of gravitational lensing.

Many sources will be lensed, which complicates parameter estimation.
However, if the lens mass is low, it will not affect the waveform in a visible manner. Similarly, if the lens is far from the line of sight it will not affect the signal. We find that events that are likely to produce detectable evidence of lensing will be in the transition region reaching moderate values for $y$ and high enough values for $M_L$. 

We use the point-mass lens model to estimate the detectability of lensed binary black hole events by a gravitational wave detector. In this simplest model for microlensing, the lensed waveform in the frequency domain is obtained by multiplying the unlensed waveform by a transmission factor. In general, the transmission factor depends on the mass of the lens  $M_L$ and on the distance from the line of sight $y$ between the source and the observer. Lensing induces (1) frequency dependent amplification and (2) distortion of the waveform arising from constructive and destructive interference between two virtual images of the source. If diffraction dominates, only the frequency dependent amplification occurs, which solely depends on $M_L$. Conversely, when two virtual images of the source interfere, the geometrical optics limit assures that the transmission factor depends only on $y$. Gravitational lensing then introduces a regular beating pattern (areas of constructive and destructive interference) that is predicted analytically.

We map the SNR increase as a function of lens parameters $y$ and $M_L$. To estimate detectability, we produce a map of the optimal match between the lensed waveform and its unlensed counterparts, while optimizing over the amplitude and phase of the gravitational wave. A mismatch of 10\% is assumed to be confidently detectable. While a mismatch of 5\% results is considered to be mild evidence for lensing. {   Lower mismatch can be relevant for high SNR cases. }

Most importantly, we show that the mismatch causes a selection bias {   that enhances smaller values of the lensing angle $y$. This modifies our expectation of the likely parameters of a first detection of lensing. This effect must be accounted for in simulations of lensing. We also show that the Bayesian evidence is diluted by including large values of $y$ in the prior, and suggest restricting to those with significant lensing.}

This paper incorporates the requirement that the lensing not only amplify the signal, but also leave a detectable imprint that confirms the existence of the microlense. This is only a fraction of the parameter space; otherwise we have 'silent' lensing. In that case the distance to the binary and the mass of the binary could be affected by lensing without creating a detectable mismatch with unlensed templates. Because the lensing does not affect these waveforms, the likelihood of the lensed and unlensed signals will be identical, and the Bayesian evidence ratio would only reflect the relative volume of the priors. In this first study we do not include a specific model that predicts the number and masses of lenses since they are so uncertain. We leave that to future work.  Instead we show the priors when they are conditioned as well on the lensed waveform having a sufficient mismatch that it leaves a measurable imprint on the signal.

Ultimately, we find that lensing  can bias the redshift distribution and that most lensed sources have low mismatch (mismatch $< 5\%$) that cannot be detected.  Furthermore, when including all lensed sources, we do not reach a redshift beyond 4 with LVK detectors. Mild evidence for lensing (mismatch of 5\% or higher) can be obtained up $z=2$ for more massive detections $M_{det}=120$, which in the source frame would be consistent with a total binary mass $M_{src}=M_1+ M_2 = 40 M_\odot$.   

We find that compact lenses (e.g., primordial black holes) of $M_L = 30 M_\odot$ in the LVK band provide only slight amplification of the gravitational signal of about $20\%$ without distortion. We conclude that in the point mass lens approximation, lenses of $20-30 M_\odot$ have a minimal effect on the detected mass of binary black holes unless they can be embedded in heavier dark matter structures. More work is needed to go beyond the point mass lens model. 

In this paper we have taken the critical density of dark matter as a reference and upper limit for the number density of lenses $n_L m_L< \rho_{cr}^{DM}$.  We note that this manuscript takes the co-moving density to be constant for both the sources and the lenses, which is a simplified assumption that is valid only when the number of mergers is negligible.
We will relax this assumption and consider more realistic distributions in future work. 

\section*{Acknowledgements}
\label{Acknowledgements}
The authors are grateful to the members of the Gravitational Waves research group from ICC UB for their continual advice and support. We particularly thank Mark Gieles, Tomas Andrade, Juan Trenado, and Daniel Mar\'{i}n Pina. We also thank ICC secretaries, Esther Pallarés Guimerà and Anna Argudo who go well beyond their job to keep everything running smoothly. We are also grateful to Tom Collett and David Bacon from ICG Portsmouth for numerous helpful discussions.

We acknowledge support from the Spanish Ministry of Science and Innovation through grant PID2021-125485NB-C22 and  CEX2019-000918-M funded by MCIN/AEI/10.13039/501100011033. RB and OB acknowledge support from the AGAUR through grant SGR-2021-01069. AL is supported by STFC grants ST/T000550/1 and ST/V005715/1.  This work has used the PyCBC python package, https://doi.org/10.5281/zenodo.7547919. The authors are grateful for computational resources provided by the LIGO Laboratory and Cardiff University and supported by National Science Foundation Grants PHY-0757058 and PHY-0823459 and STFC grant ST/I006285/1. Supporting research data are available on reasonable request from AL. For the purpose of open access, the author(s) has applied a Creative Commons Attribution (CC BY) licence to any Author Accepted Manuscript version arising.

\section*{Appendix A: the Transmission Factor in the Amplification and GO Region with Minima and Maxima}
\label{appendixA}
  {The transmission factor $F$ has a different behaviour depending on its arguments: $M_L$, $y$ and $f$. Since $F$ impacts the waveform as given by Eq.~\eqref{lensed_strain}, it is important to understand the extent of each of these behaviours depending on the different parameters.}

  {Here we briefly explain the transmission factor limits and their ranges of validity. We also analyze the predictions for its amplification, as well as the frequencies where the interference pattern has maxima and minima, which can have implications in lifting the degeneracy, e.g., between lensed  and precessing sources.}

The absolute value of the transmission factor is obtained from Eq.~\eqref{F_PML} as follows \cite{deguchi86a}
\begin{eqnarray}
|F|=\displaystyle{
\left( \frac{2 \pi^2 \nu}{1-e^{-2 \pi^2 \nu}} \right)^{1/2}
 \left| _1F_1 ( \ii \pi \nu; \,1; \,\ii \pi \nu y^2 ) \right| }.
\label{eq:abs_F}
\end{eqnarray}
Its behavior has already been described in the literature \cite{deguchi86a,takahashi03,matsunaga06} 
(in the current notations, see also Ref.~\cite{BU-22}).
Here, we briefly review the main features which will be necessary for further analysis.

The transmission factor starts from $|F|\approx 1$ at small $\nu$. Then, for a fixed source position $y$, it grows monotonically as a function of frequency before reaching its first maximum (see Fig.~\ref{fig:regions}).
This is the {\em amplification region} for which wave optics dominates.
In this region, the transmission factor is independent of $y$, as can be seen from Fig.~\ref{fig:regions}, 
and it matches an asymptotic formula \cite{deguchi86a}
\begin{align}
|F|_{\rm amp}
&=  \displaystyle{
\left( \frac{2 \pi^2 \nu}{1-e^{-2 \pi^2 \nu}} \right)^{1/2}}. 
\label{abs_Famp_exact}
\end{align}
A further approximation can be made for frequencies $\nu \gtrsim 0.3$, but still before the first maximum of $|F|$. In this case
\begin{align}
|F|_{\rm amp} \approx \, \,  (2 \pi^2 \nu)^{1/2}. 
\label{abs_Famp_approx}
\end{align}
We will use the simplicity of this expression to derive a scaling law for the SNR in the amplification region. This approximation is valid because ground-based detectors are not sensitive to low frequencies and so this form is effectively equivalent to the one above.

The extent of the amplification region depends on the lens mass: the lower $M_L$, the higher the frequency when the signal starts to be magnified due to lensing.
On the other hand, the level of magnification is determined by the lens alignment $y$.

For higher frequencies, the transmission factor starts to oscillate (see Fig.~\ref{fig:regions}) approaching the {\em geometrical optics} (GO) region, where the dominant contribution comes from two well-defined images of the source. 
In this limit, the transmission factor is given by
\begin{equation}
|F|_{\rm GO} = \left(\frac{y^2+4 \cos^2\alpha}{y \sqrt{y^2+4 }} \right)^{1/2},
\label{F_GO_amp1}
\end{equation}
where $\alpha = \pi f \Delta t_{21} - \pi/4$.
It oscillates between regular, predictable maxima and minima that correspond to constructive and destructive interference caused by the time delay $\Delta t_{21}$ between the two images and an additional Morse (topological) phase shift. 
For the ``close alignment" condition ($y \lesssim 0.5$), $\Delta t_{21}\approx 2 y\, t_M$ and 
the position of each oscillation occurs at known frequencies \cite{BU-22}
\begin{equation}
f_n = \displaystyle{
\Delta f 
\cdot
\begin{cases}
\left( n+\frac{1}{4} \right), \quad \text{at maxima,} \\[1mm]
\left( n+\frac{3}{4} \right),  \quad \text{at minima}.
\end{cases}
}
\label{eq:hyperbolas}
\end{equation}
with $n = 0, 1, 2 ...$ and the frequency spacing
\begin{equation}
\Delta f = \frac{1}{2\, t_M y} 
\simeq 2.5\times 10^4 \,{\rm Hz} \,\left(\frac{M_\odot}{M_L} \right)
\,\left(\frac{1}{y} \right).
\label{eq:freq-spacing}
\end{equation}
The onset of the GO oscillations can be assigned to a threshold frequency $f_G$ between the first maximum and the first minimum, that gives $f_G\, \Delta t_{21} \approx 1/2$ \cite{BU-22}.

We emphasize that 

(i) the lensing oscillations occur at predictable frequencies and are equally spaced in frequency with $\Delta f$ 
corresponding to the inverse of the time delay $\Delta t_{21}$ between the two images 
[Eqs.~\eqref{eq:hyperbolas} and \eqref{eq:freq-spacing}];

(ii) the amplitude of maxima and minima stays constant when $y$ is fixed, which can be seen in Fig.~\ref{fig:regions} as dotted horizontal lines:
\begin{equation}
|F|_{GO}^{\rm max} = \left(\frac{\sqrt{y^2+4}}{y}\right)^{1/2}, \, \, \, \, \, \,  |F|_{GO}^{\rm min} = \left(\frac{y}{\sqrt{y^2+4}}\right)^{1/2}.
\label{F_GO_maxmin}
\end{equation}
Note, the closer the source to the line of sight (i.e. the smaller $y$), the higher the maximum amplification, $|F|_{GO}^{\rm max}$.

(iii) in the ``close alignment" regime, $y \lesssim 0.5$, the spacing of the oscillations will be given by the inverse of the product $2y \, t_M$, while the amplification of the maxima and minima will be given by $|F|_{GO}^{\rm max} \approx \sqrt{2/y + y/4}$ and $|F|_{GO}^{\rm min} \approx \sqrt{y/2}$ \cite{BU-22}. 

These remarks are important, since the effect of the oscillations on the waveform $\tilde{h}_{UL} (f)$  can be commonly mistaken by other ``mimickers" like precession \cite{hannam14,schmidt15} or eccentricity \cite{ramos-buades22}.
For these, the waveform is modulated at the source, while the lens modulates the waveform on its way to the observer. The regular spacing of the oscillations is characteristic to the lensing effect and is related to the phase difference between the two GO paths, which includes a Morse phase shift. The Morse shift was also shown to be important for the case of strong lensing where the images are widely separated and do not interfere \cite{dai17,dai20,ezquiaga20}. 
Knowing that the oscillations are equally spaced could be one way to distinguish the lensing case from the ``mimickers" when enough of the signal is detected. As a comparison, precession does not induce regularly spaced oscillations -- there the oscillations are more pronounced at low frequencies \cite{hannam14,schmidt15}.



\section*{Appendix B: SNR increase in the Amplification and GO region}

  {In this Appendix we derive the SNR increase at the different regions: (i) amplification-only and (ii) well inside GO, which we will call ``GO average", as seen below.}

At low frequency there is only amplification, following the asymptote  [Eq.~\eqref{abs_Famp_exact}], the dashed blue line. In the amplification-only region, for $\nu \gtrsim 0.3$ when Eq.~\eqref{abs_Famp_approx} holds,
the lensed waveform is approximately
\begin{equation}
h = \sqrt{2 \pi^2 \nu} h_{UL}.
\label{hamp}
\end{equation}
If we plug this in Eq.\ (\ref{eq:snr_increase}) we obtain
\begin{equation}
    \rho_{\rm rel}^2 = 2 \pi^2 t_M K_1 \propto M_L,
    \label{rhoamp}
\end{equation}
where $K_1$ is defined by Eq.\ (\ref{eq:Ka}) with $\alpha=1$. Thus we have shown that in the amplification region $\rho_{\rm rel}$ depends only on $M_L$. The numerical values for $K_1$ and $\rho_{\rm rel}$ are displayed in Fig.\ \ref{fig:momRel}(a) and (b) as a function of $M_{det}$. Fig.\ \ref{fig:momRel}(b) provides the maximum value for $\rho_{\rm rel}$ in the amplification region for the point mass lens approximation.

For higher frequencies, oscillations appear due to interference between two images of the source, in the GO limit. 

Since the SNR integrates over the frequencies, the oscillations in Eq.~\eqref{F_GO_amp1} tend  to cancel out. When the number of oscillations is sufficient (approximately from the 8th maximum, the green region in Fig.~\ref{fig:3regions}), well beyond the onset of the oscillations, they asymptotically average out to 
\begin{equation}
 \langle |F| _{\rm GO}  \rangle \approx \sqrt{\frac{y^2+2}{y\sqrt{y^2+4}}} 
\label{eq:rho_asympt}
\end{equation}
which is independent on $f$ and $M_L$. 
Thus, in this ``GO average" region, the SNR increase depends only on $y$, as seen in Fig.~\ref{fig:snr-transition}:
\begin{equation}
\rho_{\rm rel}^2 = \frac{y^2+2}{y\sqrt{y^2+4}} \approx \frac{1}{y},
\label{eq:GO-stability}
\end{equation}
where the last approximation is valid for the ``close alignment" condition.

\section*{Appendix C: Imprint of the Lens on the Waveform in the Frequency and Time Domain}

  {It can be helpful to visualize the gravitational lensing effect in both the time-domain $h(t)$ and frequency-domain $\tilde{h} (f)$ waveforms. Here we will describe the imprint for the different lens masses considered in the text, see how it differs from the unlensed signal, and visualize the maxima and minima predicted in Appendix A.}

\begin{figure}[!htbp]
\includegraphics[width=1.0\columnwidth]{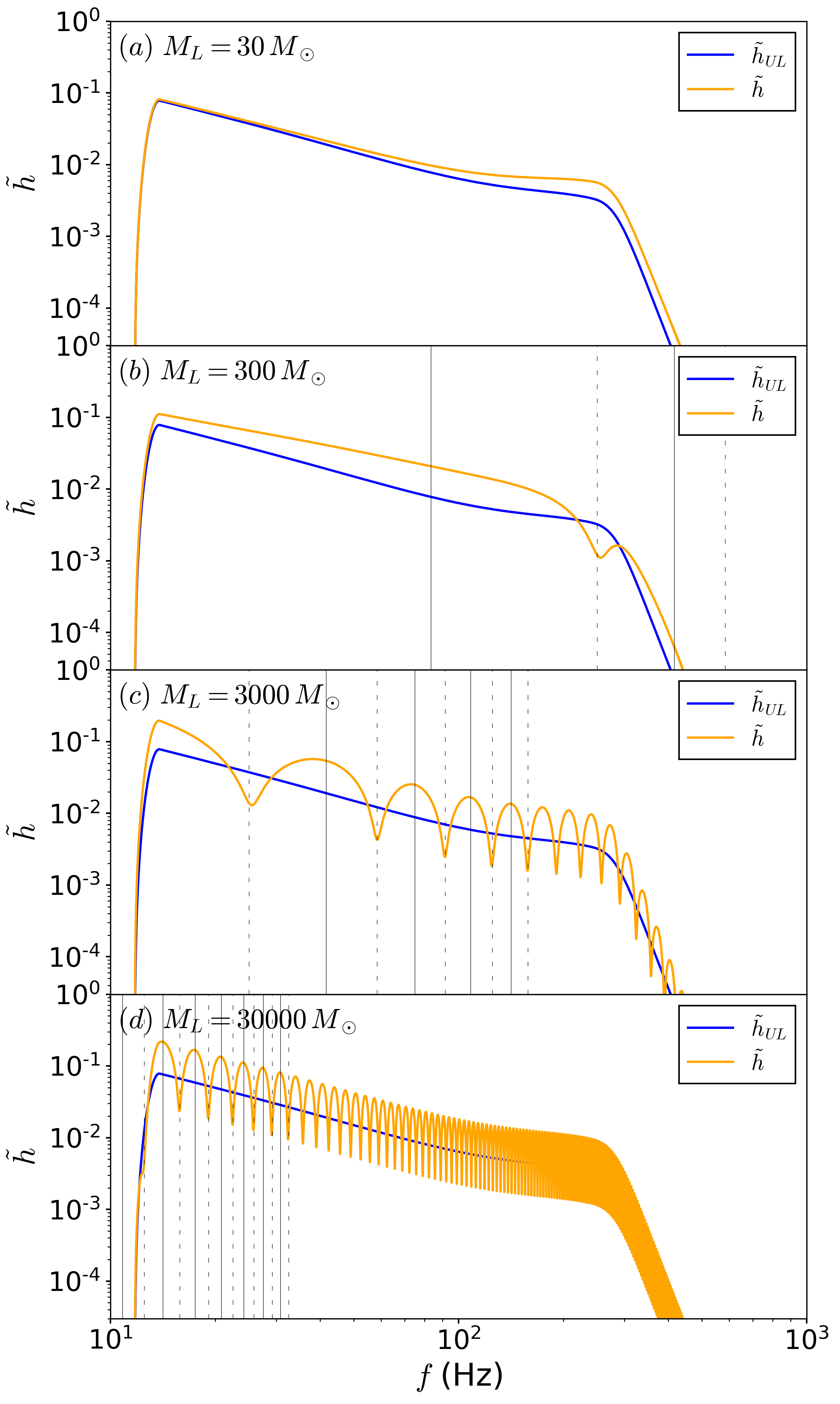}
\caption{Frequency domain strain of a lensed gravitational wave (in orange) compared to the unlensed one (in blue), as a function of frequency for $y=0.25$. The effects of lensing are seen as amplification and modulation (beating pattern, due to the interference between images). The predicted locations of the maxima and minima are shown with vertical lines. It can be seen they correspond to maxima and minima of the waveform (dashed and solid lines respectively).}
\label{fig:h-f}
\end{figure} 
\begin{figure}[!htbp]
\includegraphics[width=1.0\columnwidth]{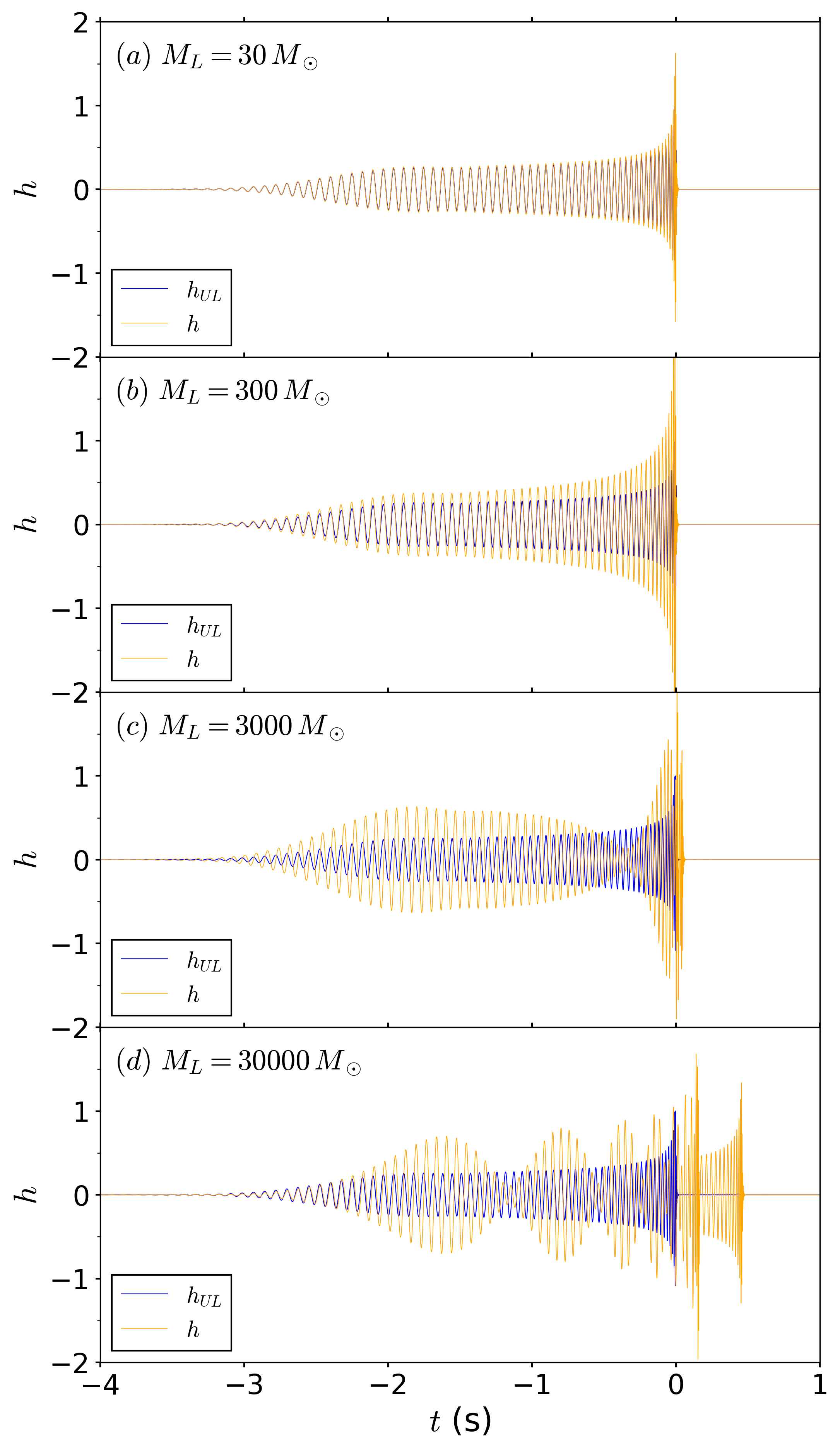}
\caption{Time domain strain of a lensed gravitational wave (in orange) compared to the unlensed one (in blue), as a function of time for $y=0.25$. The lensing effects are seen as amplification and modulation (beating pattern). When two images appear (d), both the first and the second lensed images can be seen arriving later than the unlensed one. Unlike other processes that can cause beating patterns, the presence of two separate images is a unique feature of gravitational lensing.}
\label{fig:h-t}
\end{figure} 
The transmission factor imprints the gravitational lensing effect on the original waveform $\tilde{h}_{UL} (f)$, as ${\tilde{h} (f) = \tilde{h}_{UL} (f)  F (f)}$. {It is pedagogical to visualize this imprint to understand the effect the lens has on the waveform in both in the frequency domain  
$\tilde{h} (f)$ and in its Fourier transform $h(t)$, the strain in time domain.} We consider the $M_{det} = 60 M_\odot$ gravitational waveform used in Sec.~\ref{sec_pml}.

Fig.~\ref{fig:h-f} compares the lensed $|\tilde{h} (f)|$ with the unlensed $|\tilde{h}_{UL} (f)|$ waveforms, when the position of the lens is fixed at $y=0.25$. The mass of the lens changes 
between $M_L = 30 M_\odot$ and $M_L = 30000 M_\odot$.
The two behaviours of the transmission factor can also be seen here in the imprint:

(i) When $M_L$ is small  
(i.e. for $y=0.25$, $M_L = 30 M_\odot$ in Fig.~\ref{fig:regions} and Fig.~\ref{fig:h-f}(a)),
the transmission factor $|F|$ is a monotonic function:
the amplification of the signal is gradually increasing with frequency. The highest magnification occurrs at merger (the frequency is highest there). 

(ii) For higher lens mass, 
$|F|$ has oscillations at high frequencies (from the first maximum at $f_1^+ = 1/(8 t_M y) \approx 2.5 \cdot 10^4 \, M_\odot/M_L$). The positions of constructive and destructive interference can be deduced using Eq.~\eqref{eq:hyperbolas}.
These are imprinted on the strain as seen in 
Fig.~\ref{fig:h-f}(b),(c),(d),
and are depicted with vertical solid and dashed lines respectively. 
As seen in Sec.~\ref{sec_pml}, the fringe spacing $\Delta f$ is dependent on the 
product $y \, M_L$, 
while the amplitude of the oscillations only depends on $y$.

The time evolution of the lensed gravitational wave strain, $h(t)$, is obtained by taking the Fourier transform of $\tilde{h} (f)$. The results are shown in Fig.~\ref{fig:h-t} again at fixed $y=0.25$, while $M_L$ varies. For small mass like
$M_L=30 M_\odot$
,  the lensed waveform is just amplified relative to the unlensed one.  The amplification is monotonic for $|F|$ and thus largest for the frequencies close to the merger. For higher $M_L$, a ``beating pattern" (amplitude modulation) appears, caused by the interference between the two images. The beating frequency increases with frequency \cite{hou21}. In
Fig.~\ref{fig:h-t}(d)
 we can see two separate images appearing, each one coming with a different time delay with respect to the unlensed case. The earliest signal is the interference between the two images (containing the beating pattern), followed by the the first image of the merger. Afterwards, the second image arrives alone without interference, therefore having the shape of a single chirp (but affected by magnification). Both images have different magnifications respect to the unlensed signal, which are dependent on $y$ [Eq.~\eqref{eq:magnifications}]. 
 
 Unlike the amplification and beating pattern, which can be mimicked by other processes (precession \cite{hannam14, schmidt15}, eccentricity \cite{ramos-buades22}), the detection of the two separated images is a unique and distinct feature of gravitational lensing.

 Spectrograms in the time-frequency plane reveal the signal's power, and while traditionally not employed for parameter estimation, new methods based on neural networks have emerged for identifying signals \cite{kim22,aveiro22,andres-carcasona23,ravichandran23}. Microlensed signals, characterized by distinctive maxima and minima, could be one of the classes that the network is trained on.

 To generate the spectrograms, we apply the  Q-transform algorithm \cite{chatterji04} via the PyCBC software package to the noise-added strain signal.
 \begin{figure} [!htbp]
\includegraphics[width=1.0\columnwidth]{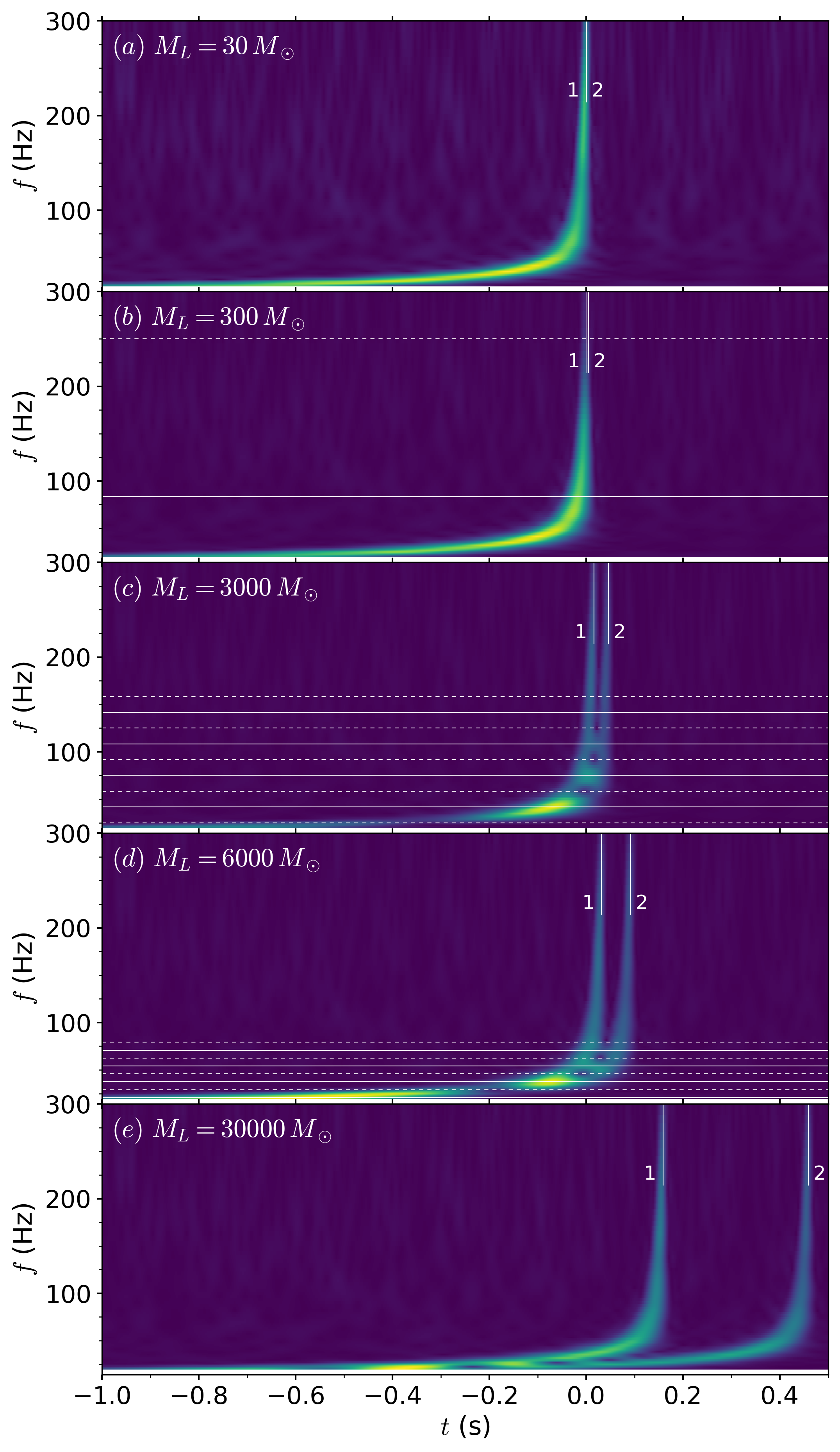}
\caption{Spectrograms of the lensed signal for $y=0.25$, and increasing $M_L$, in white noise.  In (a), the signal is amplified by 
the monotonic beginning of $|F|$.
For lower $M_L$ the interference pattern appears at higher frequencies [subplots (b) and (c)], while for higher $M_L$ it appears at lower frequencies [(subplots (d) and (e)]. The frequencies for constructive interference (maxima) and destructive interference (minima) predicted by Eq.~\eqref{eq:hyperbolas} are marked in white solid  and dashed lines respectively in subplots (b), (c), (d).  
The two images of the merger ($1$ and $2$) are marked with white vertical lines. Their separation $\Delta t_{21}$ increases with the product $y \, t_M \propto y \, M_L$.
}
\label{fig:espectrogram}
\end{figure}
Fig.~\ref{fig:espectrogram} shows the results for the parameters similar to those utilized in Fig.~\ref{fig:h-t}.
As the lens mass increases, we notice a transition from the  amplification of the original signal to the emergence of a beating pattern between two distinct images,
resembling a crab's claw. The pattern includes dim regions where the signal is suppressed and bright regions where it is enhanced. These correspond to destructive and constructive interference, respectively. 
The separation between evenly spaced maxima/minima is determined by Eq.~\eqref{eq:freq-spacing}.


Both GO images experience a time delay compared to the unlensed signal, and we can predict their positions. 
The first image (labeled as 1 in Fig.~\ref{fig:espectrogram}) is determined as the minimum of the Fermat potential \cite{schneider-92}:
$t_1=t_M\left[\frac{1}{2}(x_1-y)^2 - \ln|x_1|\right]$
with $x_1 \equiv (1/2)(y+\sqrt{y^2+4})$. 
The position of the second image (labeled as 2) is obtained by adding to $t_1$ the time delay $\Delta t_{21} \approx 2y \, t_M$ \cite{BU-22}. 
\begin{figure} [t]
\includegraphics[width=0.9\columnwidth]{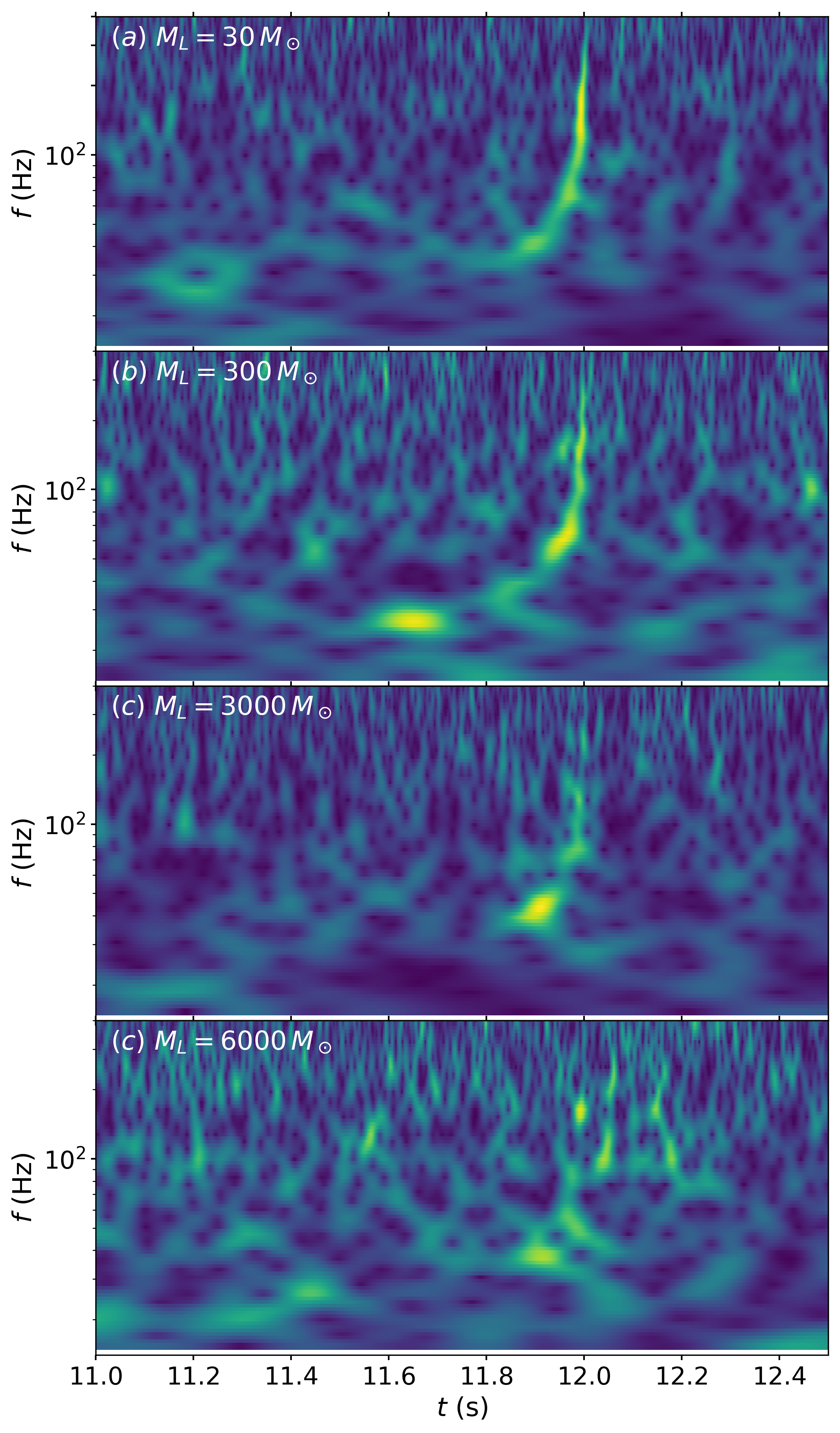}
\caption{Spectrograms of the lensed signal in simulated O4 noise at $SNR = 10$ for  $y=0.25$ and $M_L$ varying between $30 M_\odot$ and $6000 M_\odot$.}
\label{fig:espectrogram_noise2}
\end{figure} 
\begin{table}[b]
    \centering
    \begin{tabular}{|c|c|c|c|}  
 \hline
 Model & $M_L [M_\odot]$ & match & SNR Increase ($\rho_{rel}$)\\ 
 \hline
 1a &  30  & 98.5\% & 1.3 \\ 
 1b &  300  & 92.8 \% & 2.1  \\ 
 1c &  3000 & 78.4\%  & 2.0 \\ 
 1d & 6000 & 78.9 \%  & 2.0  \\
 \hline
 \end{tabular}
    \caption{Model parameters for lensed signals with  $y=0.25$ in simulated O4 noise. 
    }
    \label{tab:models}
\end{table}

{In O4 and O5, it is likely} that most signals will be detected at SNRs close to the threshold SNR. Then, in practice, the minima and maxima might not be observable by eye due to the abundance of noise.  Fig.~\ref{fig:espectrogram_noise2} shows  spectrograms in simulated O4 noise at signal-to-noise ratio $\rho= 10$ for $y=0.25$ and several lens masses. 
Some maxima and minima are visible, but the whole waveform becomes difficult to distinguish. Table \ref{tab:models} shows the match and SNR increase for Model 1a$-$d. Beyond, $M_L = 3000 M_\odot$, $\rho_{rel}$ and the mismatch stop growing since lensing is independent of $M_L$ once the GO average is reached. An injection of higher SNR will leave Table \ref{tab:models} unchanged. At $\rho \ge 20$, the spectrograms will look similar to those in Fig.~\ref{fig:espectrogram}. 

Note that spectrograms do not retain the phase information from  the underlying signal. So any parameter constraints that are obtained only using spectrogram data will be worse than those obtained using data that includes both amplitude and phase.

\end{document}